\documentclass{llncs}
\usepackage{graphicx, subfigure, amsmath, xspace, txfonts, float, wrapfig}
\usepackage{listings,lstpseudo}
\usepackage[usenames]{color}
\usepackage{mathptmx}
\usepackage{verbatim}

\newfloat{listing}{th}{lst}
\floatname{listing}{\bf Listing}

\definecolor{Gold}{rgb}{1,0.84,0}
\newcommand{\circlenumber}[1]{%
  \begin{picture}(10,10)%
    \put(5,2.5){\circle*{11}}%
    \color{Gold}%
    \put(0,0){\makebox[9pt]{\fontfamily{phv}\fontseries{b}\selectfont\scriptsize#1}}%
    \end{picture}%
}

\begin{document}

\author{Jan David Mol \and John W. Romein}
\title{The LOFAR Beam Former: \\ Implementation and Performance Analysis}
\institute{Stichting ASTRON (Netherlands Institute for Radio Astronomy) \\ Oude Hoogeveensedijk 4, 7991 PD Dwingeloo, The Netherlands \\ \email{\{mol,romein\}@astron.nl}}
\maketitle

\begin{abstract}
Traditional radio telescopes use large, steel dishes to observe radio sources. The LOFAR radio telescope is different, and uses tens of thousands of fixed, non-movable antennas instead, a novel design that promises ground-breaking research in astronomy. The antennas observe omnidirectionally, and sky sources are observed by signal-processing techniques that combine the data from all antennas.

Another new feature of LOFAR is the elaborate use of \emph{software} to do signal processing in real time, where traditional telescopes use custom-built hardware. The use of software leads to an instrument that is inherently more flexible. However, the enormous data rate (198~Gb/s of input data) and processing requirements compel the use of a supercomputer: we use an IBM Blue Gene/P.

This paper presents a collection of new processing pipelines, collectively called the beam-forming pipelines, that greatly enhance the functionality of the telescope. Where our first pipeline could only correlate data to create sky images, the new pipelines allow the discovery of unknown pulsars, observations of known pulsars, and (in the future), to observe cosmic rays and study transient events. Unlike traditional telescopes, we can observe in hundreds of directions simultaneously. This is useful, for example, to search the sky for new pulsars. The use of software allows us to quickly add new functionality and to adapt to new insights that fully exploit the novel features and the power of our unique instrument. We also describe our optimisations to use the Blue Gene/P at very high efficiencies, maximising the effectiveness of the entire telescope. A thorough performance study identifies the limits of our system.
\end{abstract}

\section{Introduction}

The LOFAR (LOw Frequency ARray) telescope is the first of a new generation of radio telescopes. Instead of using a set of large, expensive dishes, LOFAR uses many thousands of simple antennas. Every antenna observes the full sky, and the telescope is pointed through signal-processing techniques. LOFAR's novel design allows the telescope to perform wide-angle observations as well as to observe in multiple directions simultaneously, neither of which are possible when using traditional dishes. In several ways, LOFAR will be the largest telescope in the world, and will enable ground-breaking research in several areas of astronomy and particle physics~\cite{Bruyn:02}.

Another novelty is the elaborate use of software to process the telescope data in real time. Previous generations of telescopes depended on custom-made hardware to combine data, because of the high data rates and processing requirements. The availability of sufficiently powerful supercomputers however, allow the use of software to combine telescope data, creating a more flexible and reconfigurable instrument. Because LOFAR is driven by new science, flexibility in the design is essential to explore the possibilities and limits of our telescope. 

For processing LOFAR data, we use an IBM BlueGene/P (BG/P) supercomputer. The LOFAR antennas are grouped into stations, and each station sends its data (up to 198 Gb/s for all stations) to the BG/P. Inside the BG/P, the data are processed using both real-time signal-processing routines as well as two all-to-all exchanges. The output data streams are sufficiently reduced in size to be able to stream them out of the BG/P and store them on disks in our storage cluster.

In this paper, we will present the LOFAR \emph{beam former}: a collection of software pipelines that allow the LOFAR telescope to be pointed at hundreds of sources simultaneously. A \emph{beam} consists of a 1D stream of data representing the signal from a certain area in the sky, and thus is different from a correlator, that creates 2D snapshot images of the sky. Simplified, a beam former performs a weighted addition of the input signals, while a correlator multiplies the input signals.

It is LOFAR's unique design that allows us to point at many sources at once. Traditional telescopes use dishes that have a narrow field-of-view: they are only sensitive to a small region around the source they are pointed at. LOFAR's antennas are omnidirectional. Groups of antennas (\emph{stations}) are sensitive to a wide field-of-view around the source. These views, or \emph{station beams}, are sent to the BG/P, that generates weighted additions of the station input data, called \emph{tied-array beams}. Each tied-array beam represents an offset pointing within the wide field-of-view of the stations.

The primary scientific use case driving the work presented in this paper is pulsar research~\cite{Stappers:11}. A pulsar is a rapidly rotating, highly magnetised neutron star, which emits electromagnetic radiation from its poles. Similar to the behaviour of a lighthouse, the radiation is visible to us only if one of the poles points towards the Earth, and subsequently appears to us as a very regular series of pulses, with a period as low as 1.4~ms. Pulsars are weak radio sources, and their individual pulses often do not rise above the background noise that fills our universe. Our beam former can track several pulsars at LOFAR's full observational bandwidth. Alternatively, the beam former is capable of efficiently performing sky surveys to discover new pulsars (or other radio sources) by covering the sky with hundreds of tied-array beams at a reduced observational bandwidth.

The main contributions of this paper are threefold. First, we demonstrate the power of a \emph{software\/} telescope; its flexibility allows us to add new functionality with modest effort and we show how the use of supercomputer technology enables new science in astronomy and particle physics. Second, we describe the first system which allows a telescope to be pointed in hundreds of directions. Third, we elaborately analyse the performance of our application and the effectiveness of our optimisations. 

This paper is organised as follows. First, we will describe the key characteristics of the IBM BlueGene/P supercomputer in Section \ref{Sec:bluegene}. Then, we describe LOFAR and beam forming in more detail in Section \ref{Sec:LOFAR}. Section \ref{Sec:pipelines} describes the implementation of our pipelines, followed by the performance analysis in Section \ref{Sec:performance}. We briefly discuss related work in Section \ref{Sec:related-work}, and conclude in Section \ref{Sec:conclusions}.

\section{IBM BlueGene/P}
\label{Sec:bluegene}

We use an IBM BlueGene/P (BG/P) supercomputer for the real-time processing of station data. We will describe the key features of the BG/P; more information can be found elsewhere~\cite{IBM:08}. Furthermore, we will describe how our BG/P is connected to its input and output systems.

\subsection{System Description}

Our system consists of 3 racks, with 12,480 processor cores that provide 42.4 TFLOPS peak processing power. One chip contains four PowerPC~450 cores, running at a modest 850~MHz clock speed to reduce power consumption and to increase package density. Each core has two floating-point units (FPUs) that provide support for operations on complex numbers. The chips are organised in \emph{psets}, each of which consists of 64 cores for computation (\emph{compute cores}) and one chip for communication (\emph{I/O node}). Each compute core runs a fast, simple, single-process kernel,  and has access to 512 MiB of memory. The I/O nodes consist of the same hardware as the compute nodes, but additionally have a 10~Gb/s Ethernet interface connected. They run Linux, which allows the I/O nodes to do full multitasking. One rack contains 64 psets, which is equal to 4096 compute cores and 64 I/O nodes.

The BG/P contains several networks. A fast \emph{3-dimensional torus\/} connects all compute nodes and is used for point-to-point and all-to-all communications over 3.4~Gb/s links. The torus uses DMA to offload the CPUs and allows asynchronous communication. The \emph{collective network\/} is used for communication within a pset between an I/O node and the compute nodes, using 6.8~Gb/s links. In both networks, data is routed through compute nodes using a shortest path.

\subsection{External I/O}
\label{Sec:Networks}

We customised the I/O node software stack~\cite{Yoshii:10} and run a multi-threaded program on each I/O node that is responsible for the handling of both the input and the output. Unfortunately, the I/O nodes cannot saturate their 10~Gb/s Ethernet interfaces, because the 850~MHz cores do not have enough computational power to handle the overhead caused by IRQs, IP, and UDP/TCP. An I/O node can output at most 3.1~Gb/s, unless it has to handle station input (3.1~Gb/s per station), in which case it can output at most 1.1~Gb/s. We implemented a low-overhead communication protocol called FCNP~\cite{Romein:09a} to efficiently transport data between the I/O nodes and the compute nodes. The compute nodes perform the signal processing. The I/O nodes forward the results to our storage cluster, which can sustain a throughput up to 80~Gb/s.

\section{LOFAR and Beam Forming}
\label{Sec:LOFAR}

\begin{figure}[t]
\begin{minipage}[b]{40mm}
   \includegraphics[width=\textwidth]{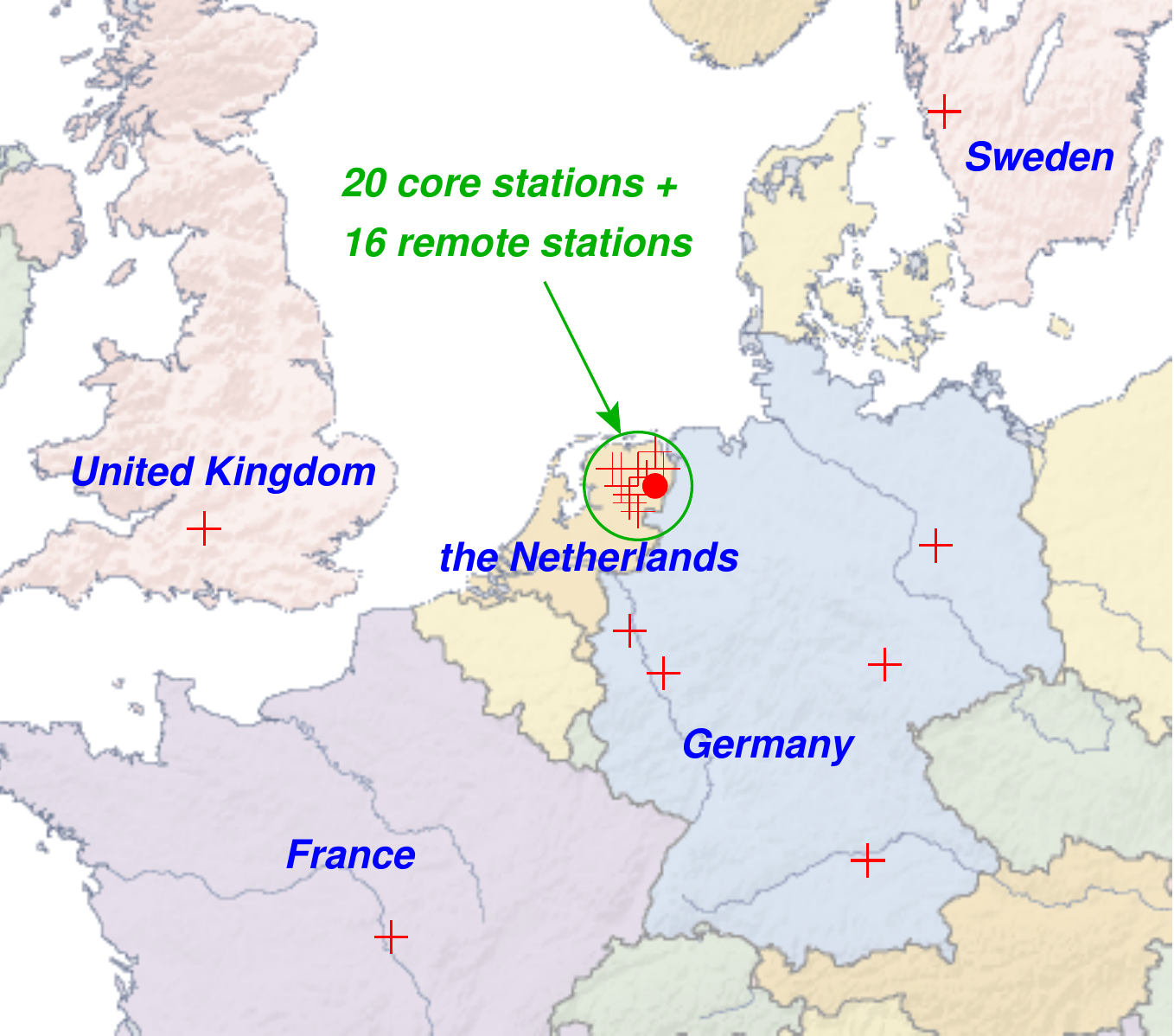}
   \caption{Locations of the stations.}
   \label{fig:map}
\end{minipage}
\hfill
\begin{minipage}[b]{35mm}
   \includegraphics[width=\textwidth]{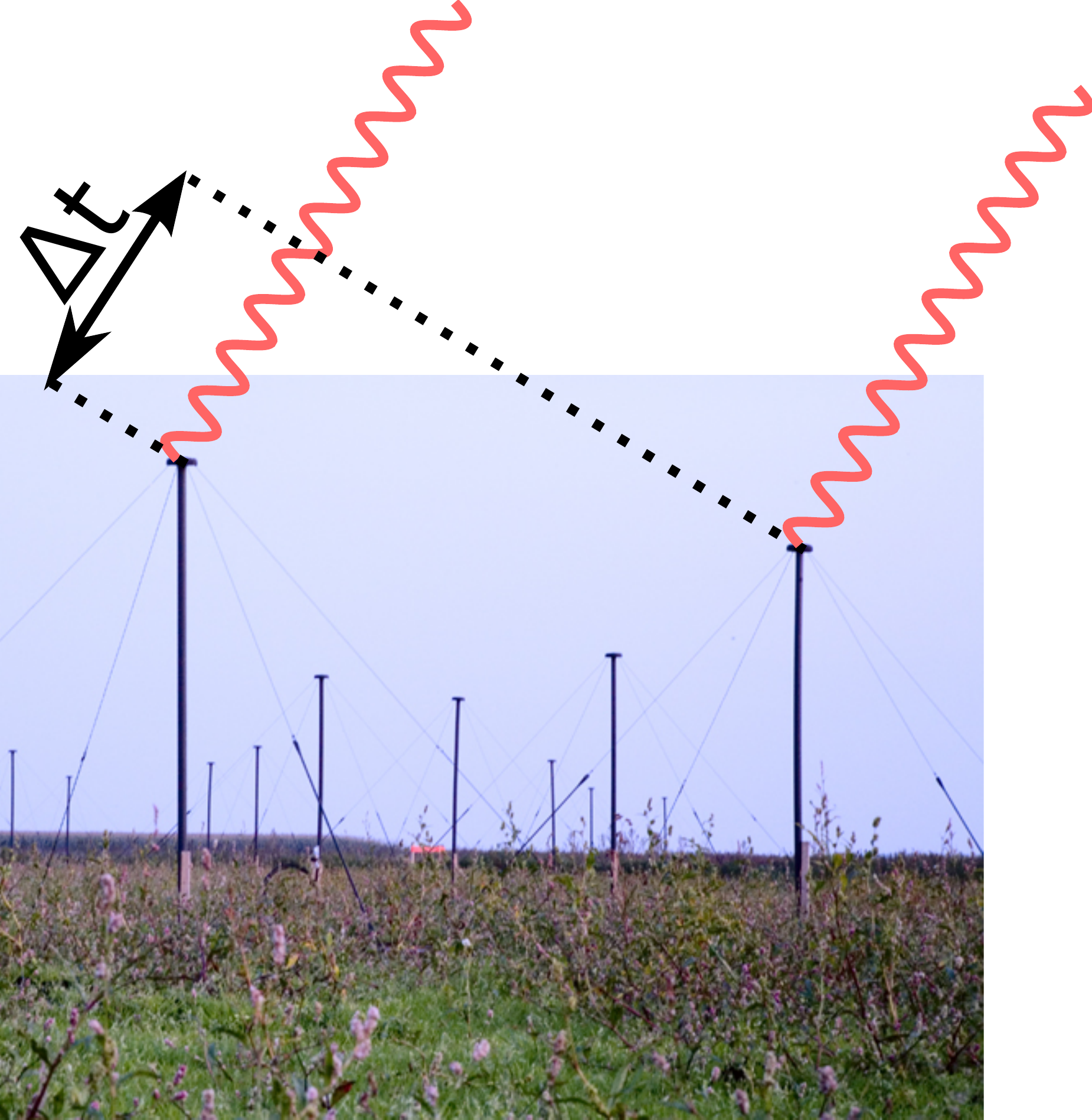}
   \caption{The left antenna receives the wave later.}
   \label{fig:delay}
\end{minipage}
\hfill
\begin{minipage}[b]{40mm}
  \includegraphics[width=0.8\textwidth]{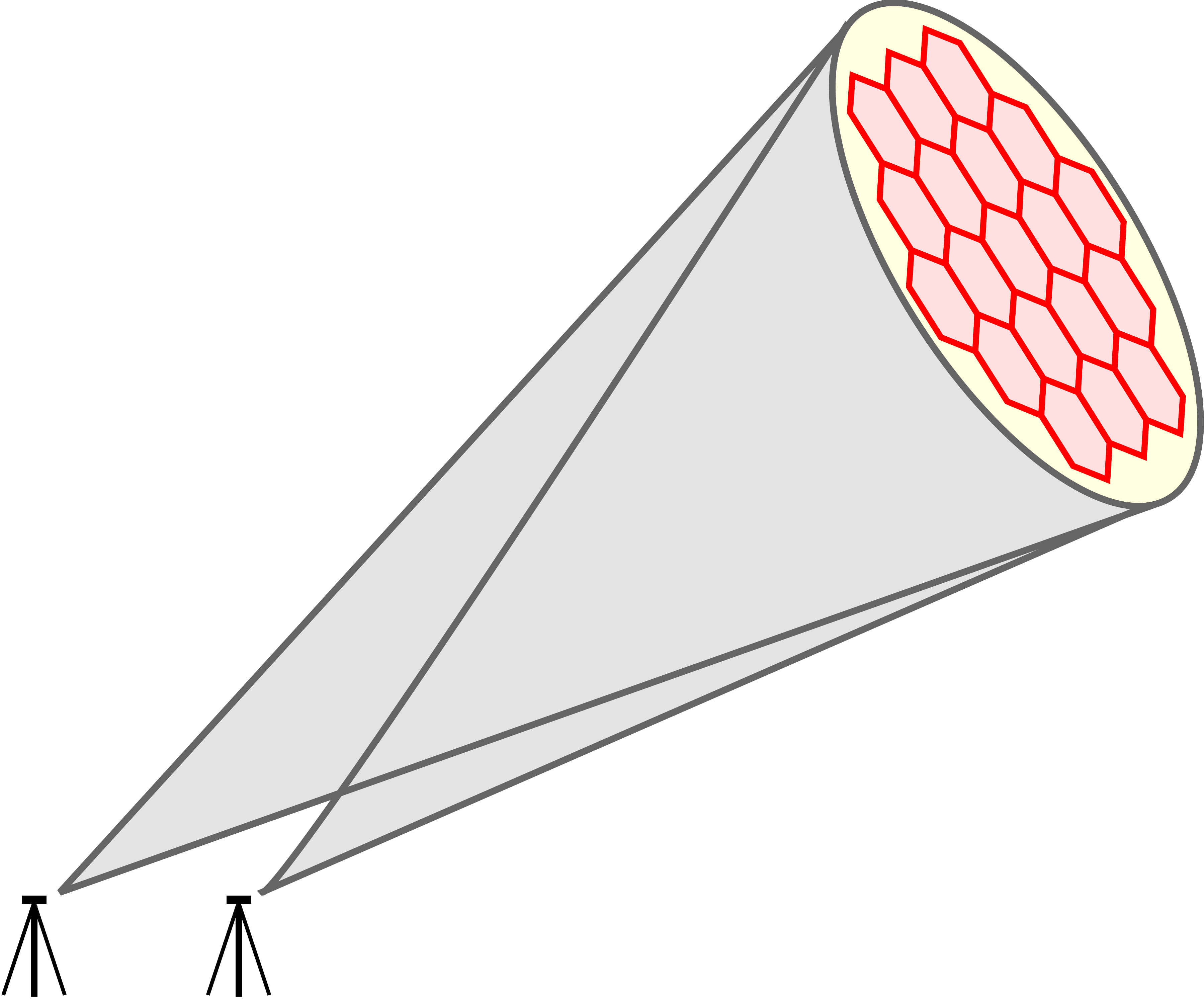}
  \caption{Tied-array beams (hexagons) formed within two station beams (ellipse).}
  \label{fig:pencilbeams}
\end{minipage}
\end{figure}

The LOFAR antennas are grouped in \emph{stations}. The stations are strategically placed, with 20 stations in the centre (the \emph{core}) and 24 stations at increasing distances from the core, spanning five nations (see Figure \ref{fig:map}). A core station can act as two individual stations in some observational modes, resulting in a total of 64 stations. A station is able to produce 248 frequency subbands of 195~kHz in the 10 -- 250~MHz sensitivity range. Each sample consists of two complex 16-bit integers, representing the amplitude and phase of the X and Y polarisations of the antennas.

Even though the antennas are omnidirectional, they can be pointed due to the fact that the speed of electromagnetic waves is finite. Signals emitted by a source reach different antennas at different times (see Figure \ref{fig:delay}). A process called \emph{delay compensation} delays the signals such that they align (are \emph{coherent}) for the desired source. Beam forming subsequently adds the aligned signals. The stations perform delay compensation and beam forming to combine the antenna signals into a station beam with a wide field-of-view. The BG/P subsequently combines the signals from different stations to form tied-array beams within the sensitive area of the station beams (see Figure \ref{fig:pencilbeams}). In the BG/P, the samples from different stations are shifted with respect to each other to compensate delay at a sample-level granularity. Sub-sample delay compensation is performed by a complex multiplication per sample, which shifts the phase of each sample. The weights used in the complex multiplication depend on the location of the stations, the observational frequency of the sample, and the sky coordinates of the tied-array beam. The beam former thus creates tied-array beams by adding the station signals using different complex weights for each beam.

Our beam former supports several pipelines. The \emph{complex voltages} pipeline stores the tied-array beams as is (X and Y polarisation samples). The \emph{Stokes IQUV} pipeline transforms the complex voltages into Stokes parameters, which are a different representation of the signal. Finally, the \emph{Stokes I} pipeline stores just the signal strength for each beam, and can be integrated in time to reduce the output data rate and to increase the number of tied-array beams that can be formed. Finally, our software can produce the Stokes parameters of an \emph{incoherent} beam, which is an accumulation of unweighted station signals. The incoherent beam is less sensitive than a coherent beam, but it maintains the wide field-of-view of the stations. The incoherent beam is typically formed in parallel with other pipelines, and is used to detect the presence of pulsars, but does not reveal their location within the station beams.

\section{Beam Former Pipelines}
\label{Sec:pipelines}

\begin{figure}[ht]
\center
\includegraphics[width=0.9\textwidth]{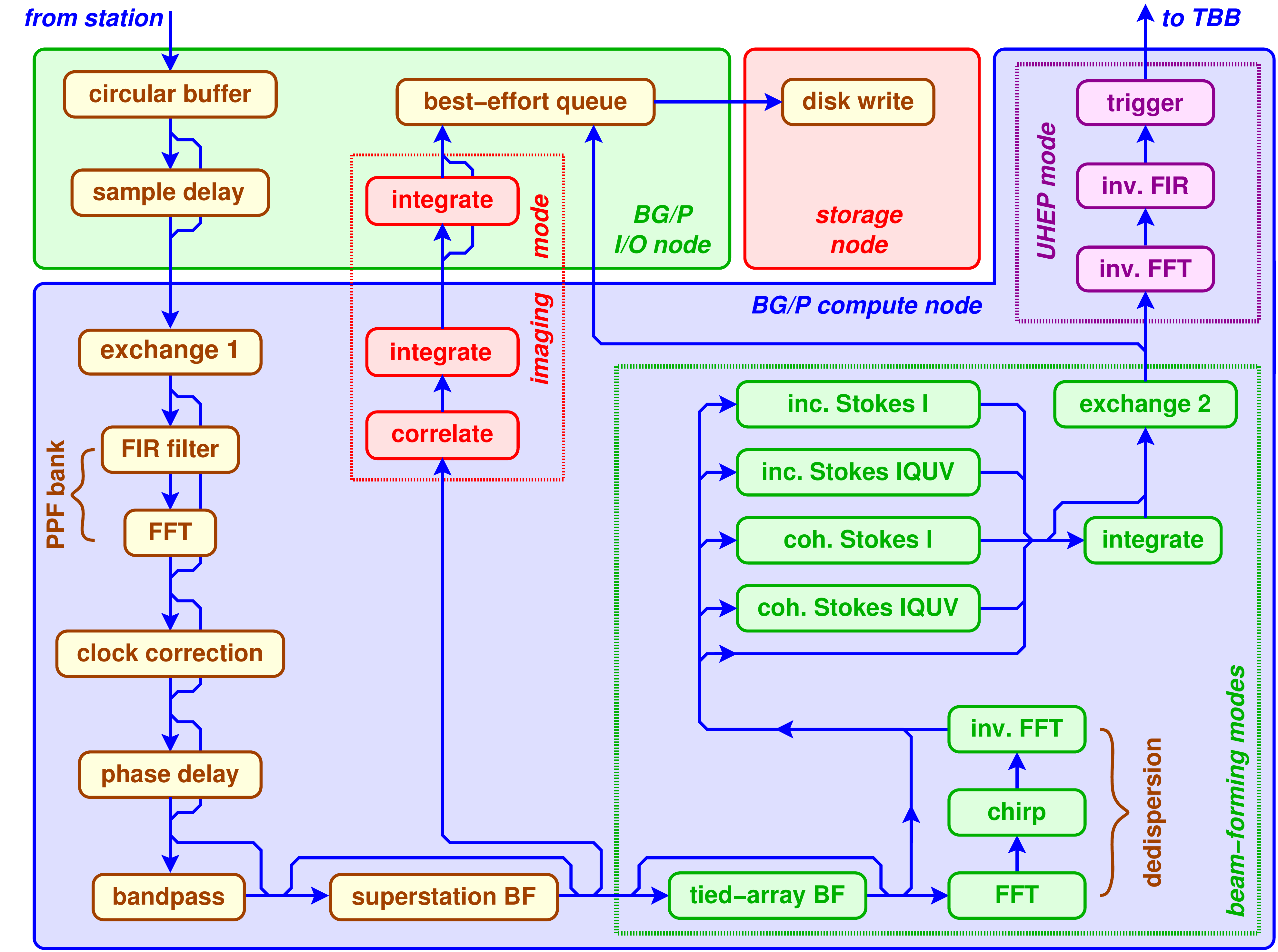}
\caption{The on-line pipelines of LOFAR. The imaging and UHEP pipelines are outside the scope of this work.}
\label{fig:processing}
\end{figure}

In this section, we will describe in detail how the full signal-processing pipelines operate, in and around the beam former. The use of a software pipeline allows us to reconfigure the components and design of our standard imaging pipeline, described elsewhere~\cite{Romein:10a}. Due to the flexibility of software, we can run several pipelines in parallel on the same data, as long as resource limits are not exceeded. Figure \ref{fig:processing} gives an overview of our system. Our software is written in C++, with core routines ported to assembly to obtain maximal performance.

\subsection{Input from Stations}
\begin{figure}[ht]
\center
\includegraphics[width=0.8\textwidth]{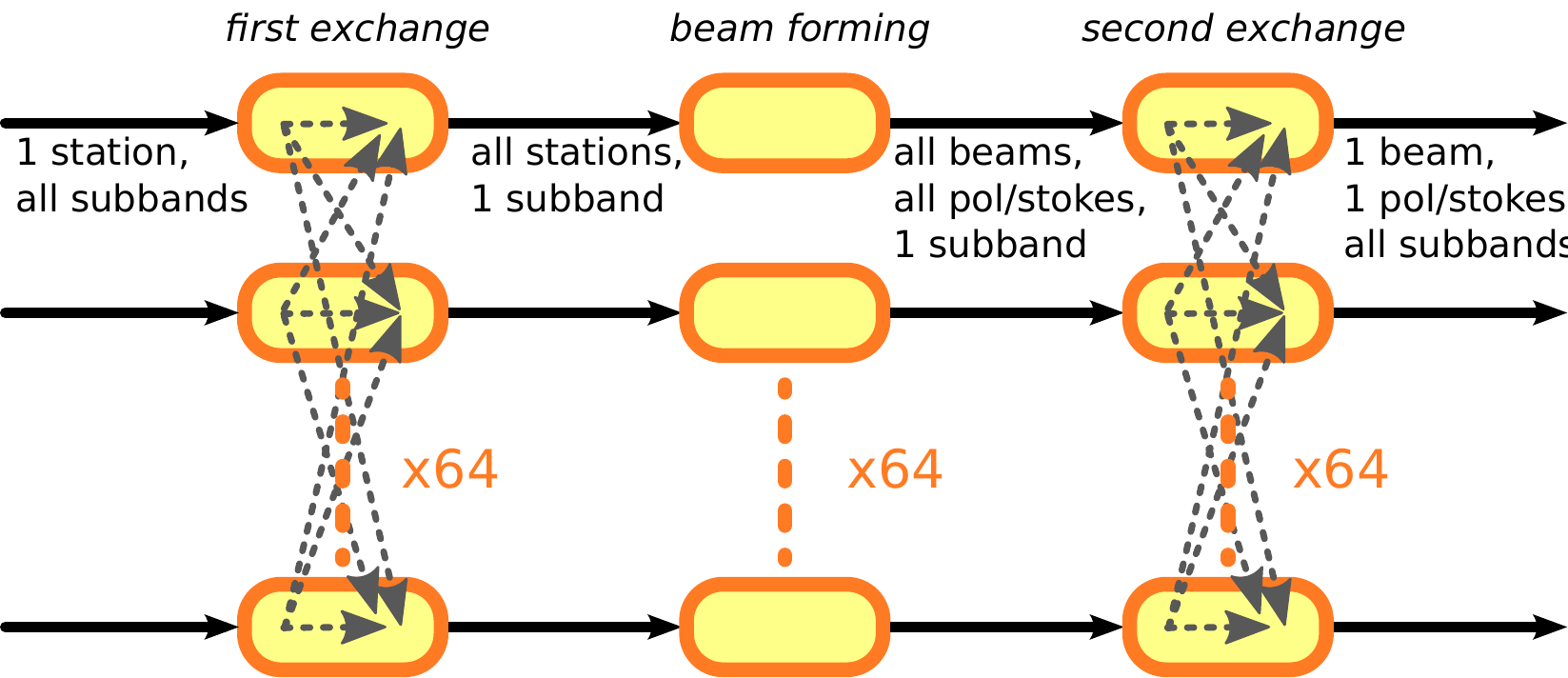}
\caption{The data flow and data ordening in our pipelines.}
\label{fig:dataflow}
\end{figure}

Each station sends data to a different I/O node. The beam former, however, needs data from all stations together to form tied-array beams. The station data thus have to be rearranged inside the BG/P, to collect the data from different stations but also to split it along different dimensions in order to distribute the workload. At the I/O nodes, the station data are split into chunks of one subband and 0.25 seconds. The chunk size is chosen such that the compute cores have enough memory to perform all of the necessary processing. Due to the BG/P design, an I/O node sends chunks to its own compute cores using the collective network. The compute cores then exchange these chunks over the torus network using an all-to-all exchange, shown in Figure \ref{fig:dataflow}.

\subsection{First All-to-all Exchange}
\label{Sec:transpose1}

The first all-to-all exchange allows the compute cores to distribute the chunks from a single station, and to collect all the chunks of the same subband from all of the stations. The exchange is performed over the fast torus network, but with up to 198~Gb/s of station data to be exchanged, special care still has to be taken to avoid network bottlenecks. It is impossible to optimise for short network paths due to the physical distances between the different psets across a BG/P rack. Instead, we optimised the data exchange by creating as many paths as possible between compute cores that have to exchange data. Within each pset, we employ a virtual mapping such that the number of possible routes between communicating cores in different psets is maximised.

The all-to-all exchange is asynchronous. Once a compute core receives a complete chunk from a single subband, it performs a sequence of processing steps on it. The first step is a conversion from 16-bit little-endian integers into 32-bit big-endian floats, to be able to use the BlueGene's powerful FPUs. Figure \ref{fig:processing} shows which steps are performed before the tied-array beam forming occurs. Note the Fast Fourier Transform (FFT) that divides the 195~kHz subbands into (typically) 12~kHz channels. We use the efficient \emph{Vienna} version of FFTW~\cite{Lorenz:05}. The superstation beam former is a simplified version of our beam former, used to combine multiple stations as if it were one, and is used in our imaging pipeline to reduce the workload. Once the chunks from all stations are received and processed asynchronously, the processed data are ready to be beam formed.

\subsection{Beam Forming}
\label{Sec:pipeline_beamforming}

The beam former combines the chunks from all stations, producing a chunk for each tied-array beam. Each beam is formed using different complex weights for the frequency of the channel, the locations of the stations, and the beam coordinates. The positional weights are precomputed by the I/O nodes and sent along with the data to avoid a duplicated effort by the compute nodes. The delays are applied to the station data through complex multiplications and additions.


All time-consuming pipeline components are written in assembly, to achieve maximum performance.  The assembly code minimises the number of memory accesses, minimises load delays, minimises FPU pipeline stalls, and maximises instruction-level parallelism.  We learnt that optimal performance is often achieved by combining multiple iterations of a multi-dimensional loops:
\begin{verbatim}
FOR Channel IN 1 .. NrChannels DO
  FOR Station IN 1 .. NrStations STEP 6 DO
    FOR Time IN 1 .. NrTimes STEP 128 DO
      FOR Beam IN 1 .. NrBeams STEP 3 DO
        BeamForm6StationsAnd128TimesTo3BeamsAssembly(...)
\end{verbatim}
This is much more efficient than to create all beams one at a time, due to better reuse of data loaded from main memory.  Finding the most efficient way to group work is a combination of careful analysis and, unfortunately, trial-and-error. The coherent beam former achieves 86\% of the FPU peak performance, not as high as the 96\% of the correlator~\cite{Romein:10a}, but still 16 times more than the C++ reference implementation. 

\subsection{Channel-level Dedispersion}

Another major component in the pulsar-observation pipeline is real-time dedispersion.  Since light of a high frequency travels faster through the interstellar medium than light of a lower frequency, the arrival time of a pulse differs for different wave lengths. To combine data from multiple frequency channels, the channels must be aligned (shifted in time). Otherwise, the pulse will be smeared or even overlap with the next pulse, causing many details to be lost. This process, called \emph{dedispersion}, is done by post-processing software that runs after the observation has finished.  However, to observe at the lowest frequencies, or to observe fast-rotating millisecond pulsars, dedispersion must also be performed \emph{within\/} a channel, since our channels (typically 12~kHz) are too wide to ignore dispersion.

\begin{figure}[ht]
\begin{minipage}[t]{0.60\textwidth}
\center
\includegraphics[width=\textwidth]{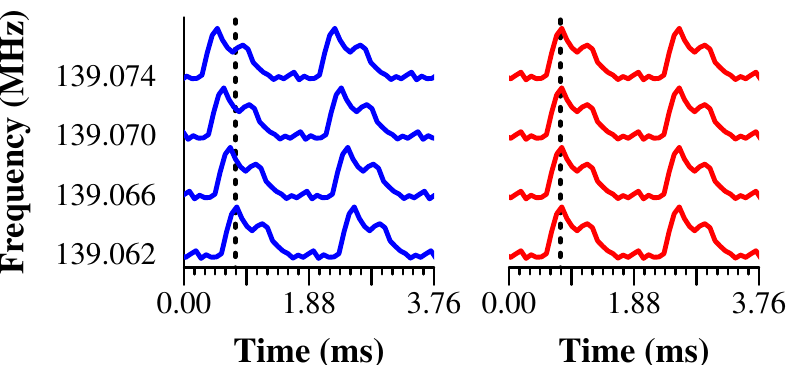}
\caption{Pulse arrival times within a 12 kHz channel before (left) and after (right) channel-level dedispersion.}
\label{fig:dispersed-signal}
\end{minipage}
\hfill
\begin{minipage}[t]{0.35\textwidth}
\center
\includegraphics[width=\textwidth]{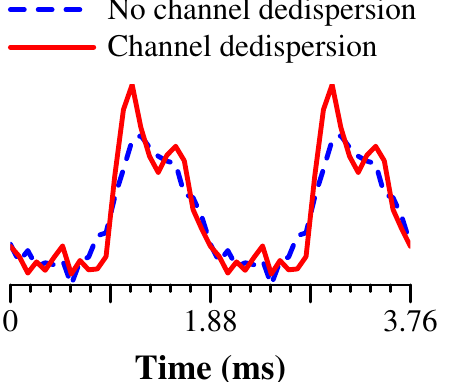}
\caption{Pulse profiles with and without channel-level dedispersion.}
\label{fig:dedispersion-result}
\end{minipage}
\end{figure}

Figure \ref{fig:dispersed-signal} shows pulses of pulsar J0034-0534 at four frequencies. The pulse period is 1.88~ms. On the left is the original dispersed signal, which results in a smeared pulse when the frequencies are collapsed into a 12~kHz channel. On the right is the dedispersed signal, which results in a sharp pulse profile when collapsed.

Dedispersion is performed in the frequency domain, effectively by doing a 4096-point FFT that splits a channel into 3~Hz subchannels. The phases of the observed samples are corrected by applying a chirp function, i.e., by multiplication with precomputed, channel-dependent, complex weights. These multiplications are programmed in assembly, to reduce the computational costs. A backward FFT is done to revert to 12~kHz channels.

Figure~\ref{fig:dedispersion-result} shows the observed effectiveness of channel-level dedispersion, which improves the effective time resolution from 0.51~ms to 0.082~ms, revealing a more detailed pulse and a better signal-to-noise ratio. Dedispersion thus contributes significantly to the data quality, but it also comes at a significant computational cost due to the two FFTs it requires. The channel-level dedispersion demonstrates the power of using a \emph{software\/} telescope: the pipeline component was implemented, verified, and optimised in only one month time.

\subsection{Stokes Calculations}

The beams are optionally converted into Stokes IQUV or Stokes I parameters, again using assembly routines to achieve optimal performance. The Stokes parameters are calculated through $I = X\overline{X} + Y\overline{Y}$, $Q = X\overline{X} - Y\overline{Y}$, $U = 2\cdot\mathrm{Re}(X\overline{Y})$, $V = 2\cdot\mathrm{Im}(X\overline{Y})$, with $\overline{X}$ as the complex conjugate of $X$. Although the formulas are simple, the Stokes parameters are expensive to calculate. The required operations for $I$ and $Q$ do not map well onto the FPU instruction set of the BG/P, even though the instruction set is extended with support for operations on complex numbers.

\subsection{Second All-to-all Exchange}

Even though the beams are formed and optionally converted into Stokes parameters, they are still distributed as chunks across the BlueGene. Because the compute nodes cannot send their data directly to the I/O node that sends it to storage, a second all-to-all exchange is required to rearrange the chunks for output. Only chunks that are sent to the same I/O node can be sent to storage as a single data stream.

Unfortunately, the output bandwidth available at each I/O node can be less than the bandwidth required by the beams. An I/O node can output 3.1~Gb/s, and only 1.1~Gb/s if the I/O node also has to process station input at the same time. The bandwidth required for a complex voltages, Stokes IQUV, or (unintegrated) Stokes I beam however is 6.2~Gb/s, 6.2~Gb/s, and 1.5~Gb/s, respectively. We therefore split the beams and send the differerent polarisations or Stokes parameters to different I/O nodes and therefore store them in different files in our storage cluster. In some cases, it is also necessary to split the beams further.

Due to memory constrains on the compute cores, the cores that performed the beam forming cannot be the same cores that receive the beam data after the second exchange. We assign a set of cores (\emph{output cores}) to receive the chunks. The output cores are chosen before an observation, and are distinct from the \emph{input cores} which perform the earlier computations in the pipeline.

The output cores again receive the chunks asynchronously, which we overlap with computations. For each chunk, the dimensions of the data are reordered into their final ordering. Reordering is necessary, because the data order that will be written to disk is not the same order that can be produced by our computations without taking heavy cache penalties. Once all of the chunks are received and reordered, they are forwarded to the I/O node.

For the distribution of the workload over the available output cores, three factors are considered. First, all of the data belonging to the same beam has to be processed by output cores in the same pset, to ensure that one I/O node can concatenate all of the 0.25 second chunks that belong to the beam. Second, the maximum output rate per I/O node has to be respected. Finally, the presence of the first all-to-all exchange, which uses the same network at up to 198~Gb/s. The second exchange uses up to 81~Gb/s. Even though each link sustains 3.4~Gb/s, it has to process the traffic from four cores, as well as traffic routed through it between other nodes. The network links in the BG/P become overloaded unless the output cores are scattered sufficiently.

\subsection{Transport to Disks}
Once an output core has received and reordered all of its data, the data are sent to the core's I/O node. The I/O node forwards the data over TCP/IP to the storage cluster. To avoid any stalling in our pipeline due to network congestion or disk issues, the I/O node uses a best-effort buffer which drops data in the unusual case that it cannot be sent.

\section{Performance Analysis}
\label{Sec:performance}

We will focus our performance analysis on the most challenging cases that are of astronomical interest. In all cases, we respect the real-time nature of our system by limiting the load such that there is at most 0.1\% of data loss, but typically, data loss is much rarer. Almost all variance occurs in the networks within the BG/P due to clashes caused by scheduling intricacies. We present measurements for a single BG/P rack.

\subsection{Overall Performance}

\begin{wrapfigure}{r}{0.5\textwidth}
\vspace{-1.65cm}
\includegraphics[width=0.5\textwidth]{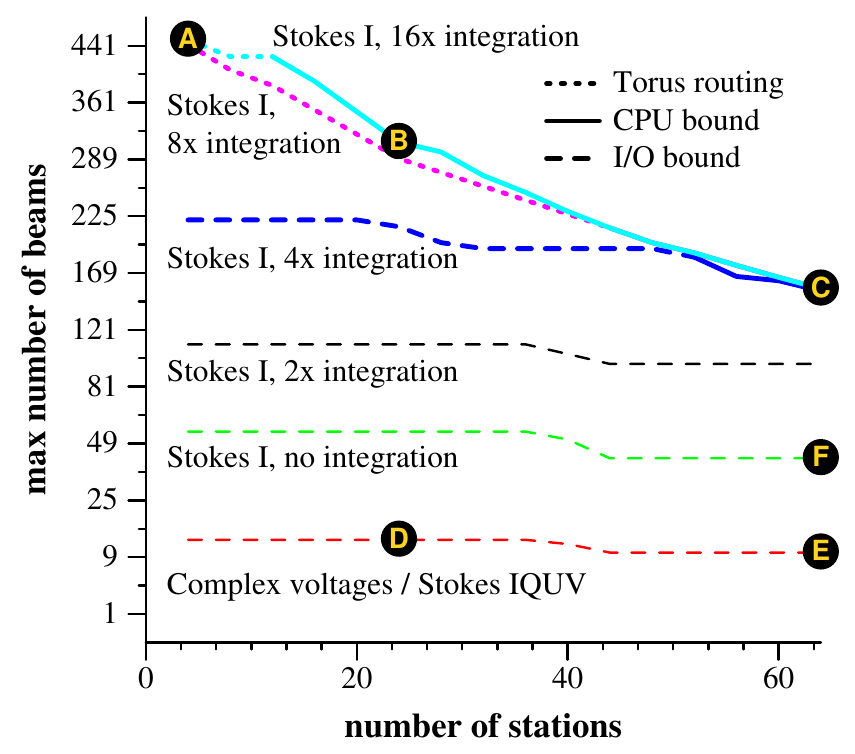}
\caption{The maximum number of beams that can be created in various configurations.}
\label{fig:stations-beams}
\vspace{-1cm}
\end{wrapfigure}

Figure \ref{fig:stations-beams} shows the maximum number of beams that can be created when using a various number of stations, in each of the three pipelines: complex voltages, Stokes IQUV, and Stokes I. Both the complex voltages and the Stokes IQUV pipelines are I/O bound. Each beam is 6.2~Gb/s wide. We can make at most 13 beams without exceeding the available 81~Gb/s to our storage cluster. If 64 stations are used, the available bandwidth is 70~Gb/s due to the fact that an I/O node can only output 1.1~Gb/s if it also has to process station data. The granularity with which the output can be distributed over the I/O nodes, as well as scheduling details, determine the actual number of beams that can be created, but in all cases, the beam former can create at least 10 beams at LOFAR's full observational bandwidth.

In the Stokes I pipeline, we applied several integration factors (1, 2, 4, 8, and 16) in order to show the trade-off between beam quality and the number of beams. Integration factors higher than 16 does not allow significantly more beams to be created, but could be used in order to further reduce the total output rate. For low integration factors, the beam former is again limited by the available output bandwidth. At 8x integration, the number of beams is limited by the virtual mapping we applied to optimise both of the all-to-all exchanges (see Section \ref{Sec:transpose1}): the high number of routes causes more collisions than the compute cores have spare time for to handle. With higher integration factors, a few more beams can be formed before the compute cores run out of computational resources. For observations for which a high integration factor is acceptable, the beam former is able to form between 155 and 450 tied-array beams, depending on the number of stations used. For observations that need a high time resolution and thus a low integration factor, the beam former is still able to form at least 42 tied-array beams.

\begin{figure}[t]
\begin{minipage}[t]{0.55\textwidth}
\includegraphics[width=\textwidth]{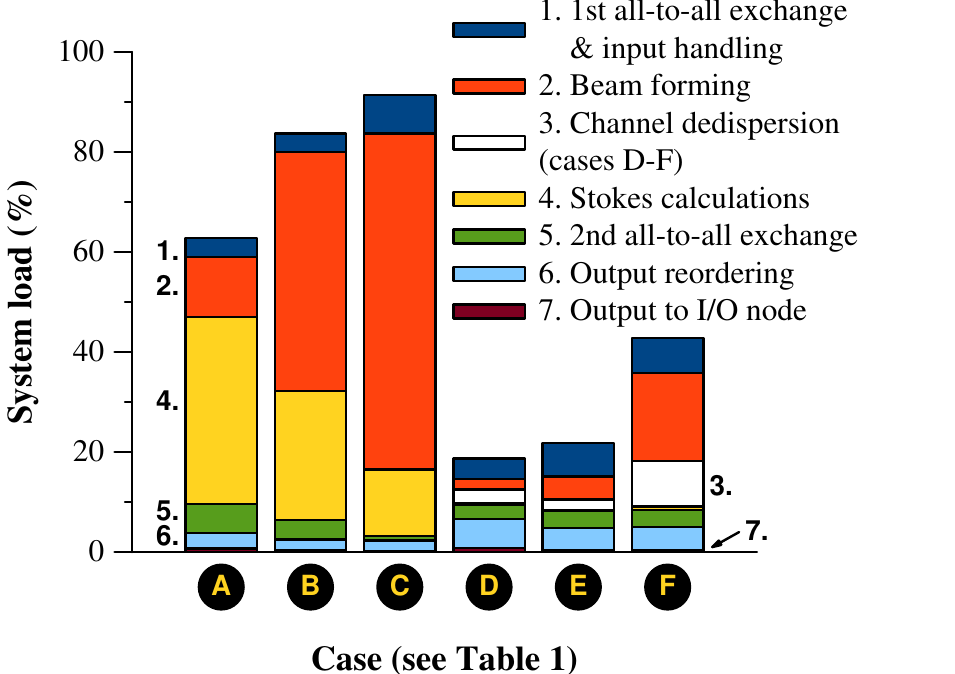}
\caption{The load of the compute cores.}
\label{fig:execution-times}
\end{minipage}
\hspace{-1cm}
\begin{minipage}[t]{0.55\textwidth}
\includegraphics[width=\textwidth]{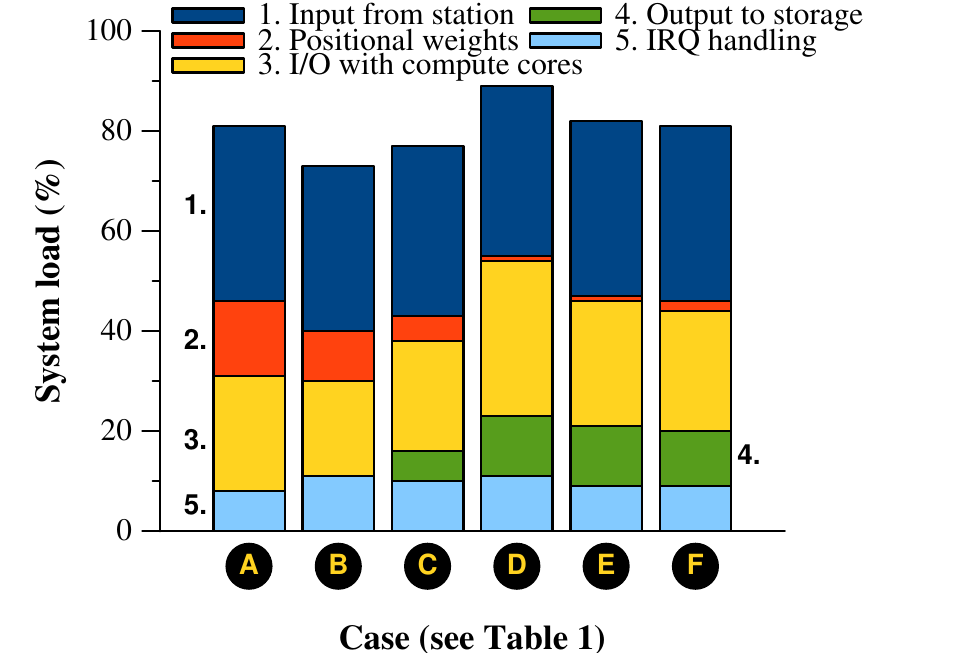}
\caption{The load of the busiest I/O nodes.}
\label{fig:ionperf}
\end{minipage}
\end{figure}

\begin{table}[t]
\caption{Several highlighted cases.}
\label{table:cases}
\center
\begin{tabular}{llrrrrrrll}
\hline\noalign{\smallskip}
Case & Mode & Channel & Int. & Stations & Beams  & Input & Output & Bound & Used for \\
     &      & dedisp. & factor      &          &        & rate  & rate   &       & \\
\noalign{\smallskip}     
\hline
\noalign{\smallskip}     
\circlenumber{A} & Stokes I    & N & 16 &  4 & 450 &  12 Gb/s & 44 Gb/s & Torus & Surveys \\
\circlenumber{B} & Stokes I    & N & 16 & 24 & 310 &  74 Gb/s & 30 Gb/s & CPU   & Surveys \\
\circlenumber{C} & Stokes I    & N &  8 & 64 & 155 & 198 Gb/s & 30 Gb/s & CPU   & Surveys \\  
\circlenumber{D} & Stokes IQUV & Y & - & 24 &  13 &  74 Gb/s & 81 Gb/s & I/O   & Known sources \\
\circlenumber{E} & Stokes IQUV & Y & - & 64 &  10 & 198 Gb/s & 62 Gb/s & I/O   & Known sources \\
\circlenumber{F} & Stokes I    & Y & 1 & 64 &  42 & 198 Gb/s & 65 Gb/s & I/O   & Known sources \\
\hline
\end{tabular}
\end{table}

\subsection{System Load}

We further analyse the workload of the compute cores by highlighting a set of cases, summarised in Table \ref{table:cases}. We will focus on case \circlenumber{A}, which creates the highest number of beams, and on CPU-bound cases useful for performing surveys, with either 24 stations (\circlenumber{B}) or 64 stations (\circlenumber{C}) as input. Cases \circlenumber{D} and \circlenumber{E} represent high-resolution observations of known sources, and are I/O bound configurations with 24 and 64 stations, respectively. Case \circlenumber{F} focusses on the observations of known sources as well, using Stokes I output, which allows more beams to be created. Channel-level dedispersion is applied for all cases that observe known sources.

The average workload of the compute cores for each case is shown in Figure \ref{fig:execution-times}. For the CPU-bound cases \circlenumber{B} and \circlenumber{C}, the average load has to be lower than 100\% to recover from small delays in the processing, that can occur since the BG/P is not a real-time system. These fluctuations typically occur due to clashes within the BG/P torus network which is used for both all-to-all-exchanges, and cannot be avoided in all cases.

In the cases where we create many beams (\circlenumber{A}\circlenumber{B}\circlenumber{C}), most of the cycles are spent on beam forming and on calculating the Stokes I parameters. The beam forming scales with both the number of stations and the number of beams, while the Stokes I calculation costs depends solely on the number of beams. Case \circlenumber{A} has to beam form only four stations, and thus requires most of its time calculating the Stokes I parameters. Cases \circlenumber{B} and \circlenumber{C} use more stations, and thus need more time to beam form.

The costs for both the first and the second all-to-all exchange are mostly hidden due to overlaps with computation. The remaining cost for the second exchange is proportional to the output bandwidth required in each case.

For the I/O-bound cases \circlenumber{D}\circlenumber{E}\circlenumber{F}, only a few tied-array beams are formed and transformed into Stokes I(QUV) parameters, which produces a lot of data but requires little CPU time. Enough CPU time is therefore available to include channel-level dedispersion, which scales with the number of beams and is an expensive operation.

Figure \ref{fig:ionperf} shows the workload for the busiest I/O nodes in each case, including the system time spent to handle IRQs. The processing of station data and the communication with the compute cores cause most of the load. In cases \circlenumber{A}\circlenumber{B}, the output is handled by I/O nodes that do not process station data. In both cases, a significant amount of time is spent computing the positional weights (see Section \ref{Sec:pipeline_beamforming}). A similar amount of time is required in cases \circlenumber{C}\circlenumber{D}\circlenumber{E}\circlenumber{F}\/ to process the output.

\section{Related Work}
\label{Sec:related-work}

The LOFAR beam former is the only beam former capable of producing hundreds of tied-array beams. A radio dish can be extended to focus on multiple sources by placing additional receivers in its focal point (a \emph{focal plane array})~\cite{Staveley-Smith:96}, but such a solution does not scale. The Murchison Widefield Array (MWA) uses a design similar to LOFAR, and plans to build a beam former, but is still under construction~\cite{Lonsdale:09}.



\section{Conclusions}
\label{Sec:conclusions}

We have shown the capabilities of our beam former pipelines, running in software on an IBM BlueGene/P supercomputer. Our system is capable of producing 13 tied-array beams at LOFAR's full observational bandwidth before our output limit of 81~Gb/s is met. Alternatively, it can form hundreds of beams at a reduced resolution, the exact number depending on the number of stations and the pipeline used. Finally, an incoherent beam can be created, which retains the wide field-of-view offered by our stations. None of these feats are possible with any other telescope.

The use of a software solution on powerful interconnected hardware is a key aspect in the development and deployment of our pipeline. Because we use software, rapid prototyping is cheap, allowing novel features to be tested to aid the exploration of the design space of a new instrument. The resulting pipelines retain the flexibility that software allows. The control flow and bookkeeping has become complex while remaining manageable through software abstraction. We are able to run the same station data through multiple pipelines in parallel, and even multiple independent observations in parallel, as long as there are enough available resources. The science which drives LOFAR, and which is driven by it, is greatly accelerated through the use of an easily reconfigurable instrument.

The BG/P supercomputer provides us with enough computing power and powerful networks to be able to implement the signal processing and all-to-all-exchanges that we require, without having to resort to a dedicated system which inevitably curbs the design freedom that the supercomputer provides. As with any system, platform-specific parameters nevertheless become important when maximal performance is desired. Although a C reference implementation allowed us to quickly develop and test features, we needed handcrafted assembly to keep the double FPUs of each compute core busy.

The architecture of the BG/P makes some tasks more difficult as well. The fact that an I/O node can only communicate with its own compute cores prevents us from freely scheduling the workload. Instead, we have to manually route the data using two all-to-all exchanges in order to stream the data from and to the right I/O nodes. To achieve maximum performance, we tuned the distribution of the workload over the cores to avoid network collisions.

\bibliographystyle{plain}
\bibliography{lofar}

\end{document}